\documentclass[12pt, centerh1]{article}

\usepackage{natbib}
\usepackage{amsmath}
\usepackage{amsfonts,mathrsfs,moreverb,lipsum}
\usepackage{graphicx,colonequals}
\usepackage{fixltx2e}
\usepackage{color}
\usepackage{caption}
\usepackage{subcaption}
\usepackage{pbox}

\newcommand{\btheta}{{\boldsymbol{\theta}}}
\newcommand{\bvtheta}{{\boldsymbol{\vartheta}}}
\newcommand{\pig}{\pi_g}

\newcommand{\vecx}{\mathbf{x}}

\newcommand{\vecX}{\mathbf{X}}

\newcommand{\matsig}{\mathbf\Sigma}
\newcommand{\matPsi}{\mathbf\Psi}

\newcommand{\tr}{\,\mbox{tr}}

\newcommand{\vecA}{\mathbf{A}}

\newcommand{\matm}{\mathbf{M}}

\newcommand{\vecc}{\text{vec}}
\newcommand{\fX}{\mathscr{X}}
\newcommand{\fV}{\mathscr{V}}
\newcommand{\tgamma}{\tilde{\gamma}}
\title{Finite Mixtures of Skewed Matrix Variate Distributions}
\author{Michael P.B. Gallaugher and Paul D. McNicholas}
\date{\small Dept.\ of Mathematics \& Statistics, McMaster University, Hamilton, Ontario, Canada.}

\begin{document}
\maketitle
\begin{abstract}
Clustering is the process of finding underlying group structures in data. Although mixture model-based clustering is firmly established in the multivariate case, there is a relative paucity of work on matrix variate distributions and none for clustering with mixtures of skewed matrix variate distributions. Four finite mixtures of skewed matrix variate distributions are considered. Parameter estimation is carried out using an expectation-conditional maximization algorithm, and both simulated and real data are used for illustration.\\[-6pt]


\end{abstract}

\section{Introduction}
Over the years, there has been increased interest in the applications involving three-way (matrix variate) data.
Although there are countless examples of clustering for multivariate distributions using finite mixture models, as discussed in Section~\ref{sec:back}, there is very little work for matrix variate distributions. Moreover, the examples in the literature deal exclusively with symmetric (non-skewed) matrix variate distributions such as the matrix variate normal and the matrix variate $t$ distributions. 

There are many different areas of application for matrix variate distributions. One area is multivariate longitudinal data, where multiple variables are measured over time \citep[e.g.,][]{Anderlucci15}. In this case, each row of a matrix would correspond to a time point and the columns would represent each of the variables. Furthermore, the two scale matrices, a defining characteristic of matrix variate distributions, allow for simultaneous modelling of the inter-variable covariances as well as the temporal covariances. A second application, considered herein, is image recognition. In this case, an image is analyzed as an $n\times p$ pixel intensity matrix.
Herein, a finite mixture of four different skewed matrix distributions, the matrix variate skew-$t$, generalized hyperbolic, variance-gamma and normal inverse Gaussian (NIG) distributions are considered. These mixture models are illustrated for both clustering (unsupervised classification) and semi-supervised classification using both simulated and real data.


\section{Background}\label{sec:back}

\subsection{Model-Based Clustering and Mixture Models}\label{sec:21}
Clustering and classification look at finding and analyzing underlying group structures in data. 
One common method used for clustering is model-based, and generally makes use of a $G$-component finite mixture model. A multivariate random variable $\vecX$ from a finite mixture model has density
$$
f(\vecx~|~\bvtheta)=\sum_{g=1}^G\pig f_g(\vecx~|~\btheta_g),
$$
where $\bvtheta=\left(\pi_1,\pi_2,\ldots,\pi_G,\btheta_1,\btheta_2,\ldots,\btheta_G\right)$, $f_g(\cdot)$ is the $g$th component density, and $\pig>0$ is the $g$th mixing proportion such that $\sum_{i=1}^G\pig=1$.
\cite{mcnicholas16a} traces the association between clustering and mixture models back to \cite{tiedeman55}, and the earliest use of a finite mixture model for clustering can be found in \cite{wolfe65}, who uses a Gaussian mixture model. Other early work in this area can be found in \cite{baum70} and \cite{scott71}, and a recent review of model-based clustering is given by \cite{mcnicholas16b}.

Although the Gaussian mixture model is well-established for clustering, largely due to its mathematical tractability, quite some work has been done in the area of non-Gaussian mixtures. For example, some work has been done using symmetric component densities that parameterize concentration (tail weight), e.g., the $t$ distribution \citep{peel00,andrews11a,andrews12,lin14} and the power exponential distribution \citep{dang15}. There has also been work on mixtures for discrete data \citep[e.g.,][]{karlis07,bouguila09} as well as several examples of mixtures of skewed distributions such as the NIG distribution \citep{karlis09,subedi14}, the skew-$t$ distribution \citep{lin10,vrbik12,vrbik14,murray14a,lee14,lee16,murray14b,murray17}, the shifted asymmetric Laplace distribution \citep{morris13b,franczak14}, the variance-gamma distribution \citep{smcnicholas17}, the generalized hyperbolic distribution \citep{browne15}, and others \citep[e.g.,][]{elguebaly15,franczak15}.

Recently, there has been an interest in the mixtures of matrix variate distributions, e.g., \cite{Anderlucci15} consider multivariate longitudinal data with the matrix variate normal distribution and \cite{dougru16} consider a finite mixture of matrix variate $t$ distributions.

\subsection{Matrix Variate Distributions}
Three-way data such as multivariate longitudinal data or greyscale image data can be easily modelled using a matrix variate distribution. There are many examples of such distributions presented in the literature, the most notable being the matrix variate normal distribution. 
In Section~\ref{sec:21}, $\vecX$ was used in the typical way to denote a multivariate random variable and $\vecx$ was used to denote its realization.
Hereafter, $\vecX$ is used to denote a realization of a random matrix $\fX$. 
An $n\times p$ random matrix $\fX$ follows an $n\times p$ matrix variate normal distribution with location parameter $\matm$ and scale matrices $\matsig$ and $\matPsi$ of dimensions $n\times n$ and $p\times p$, respectively, denoted by $\mathcal{N}_{n\times p}(\matm, \matsig, \matPsi)$ if the density of $\fX$ can be written as
\begin{equation}
f(\vecX~|~\matm, \matsig, \matPsi)=\frac{1}{(2\pi)^{\frac{np}{2}}|\matsig|^{\frac{p}{2}}|\matPsi|^{\frac{n}{2}}}\exp\left\{-\frac{1}{2}\tr\left(\matsig^{-1}(\vecX-\matm)\matPsi^{-1}(\vecX-\matm)'\right)\right\}.
\end{equation}
A well known property of the matrix variate normal distribution \citep{harrar08} is 
\begin{equation}
\fX\sim \mathcal{N}_{n\times p}(\matm,\matsig,\matPsi) \iff \vecc(\fX)\sim \mathcal{N}_{np}(\vecc(\matm),\matPsi\otimes \matsig),
\label{eq:normprop}
\end{equation}
where $\mathcal{N}_{np}(\cdot)$ is the multivariate normal density with dimension $np$, $\vecc(\cdot)$ is the vectorization operator, and $\otimes$ is the Kronecker product.

The matrix variate normal distribution has many elegant mathematical properties that have made it so popular, e.g., \cite{viroli11} uses a mixture of matrix variate normal distributions for clustering. However, there are non-normal examples such as the Wishart distribution \citep{Wishart} and the skew-normal distribution, e.g., \cite{chen2005}, \cite{dominguez2007}, and \cite{harrar08}. More information on matrix variate distributions can be found in \cite{gupta99book}.

\subsection{The Generalized Inverse Gaussian Distribution} 
The generalized inverse Gaussian distribution has two different parameterizations, both of which will be useful. A random variable $Y$ has a generalized inverse Gaussian (GIG) distribution parameterized by $a, b$ and $\lambda$, denoted by $\text{GIG}(a,b,\lambda)$, if its probability density function can be written as
$$
f(y|a, b, \lambda)=\frac{\left({a}/{b}\right)^{\frac{\lambda}{2}}y^{\lambda-1}}{2K_{\lambda}(\sqrt{ab})}\exp\left\{-\frac{ay+{b}/{y}}{2}\right\},
$$
where
$$
K_{\lambda}(u)=\frac{1}{2}\int_{0}^{\infty}y^{\lambda-1}\exp\left\{-\frac{u}{2}\left(y+\frac{1}{y}\right)\right\}dy
$$
is the modified Bessel function of the third kind with index $\lambda$. 
Expectations of some functions of a GIG random variable have a mathematically tractable form, e.g.:
\begin{equation}
\mathbb{E}(Y)=\sqrt{\frac{b}{a}}\frac{K_{\lambda+1}(\sqrt{ab})}{K_{\lambda}(\sqrt{ab})},
\label{eq:ai}
\end{equation}
\begin{equation}
\mathbb{E}\left({1}/{Y}\right)=\sqrt{\frac{a}{b}}\frac{K_{\lambda+1}(\sqrt{ab})}{K_{\lambda}(\sqrt{ab})}-\frac{2\lambda}{b},
\label{eq:bi}
\end{equation}
\begin{equation}
\mathbb{E}(\log Y)=\log\left(\sqrt{\frac{b}{a}}\right)+\frac{1}{K_{\lambda}(\sqrt{ab})}\frac{\partial}{\partial \lambda}K_{\lambda}(\sqrt{ab}).
\label{eq:ci}
\end{equation}

Although this parameterization of the GIG distribution will be useful for parameter estimation, for the purposes of deriving the density of the matrix variate generalized hyperbolic distribution, it is more useful to take the parameterization
\begin{equation}
g(y|\omega,\eta,\lambda)= \frac{\left({w}/{\eta}\right)^{\lambda-1}}{2\eta K_{\lambda}(\omega)}\exp\left\{-\frac{\omega}{2}\left(\frac{w}{\eta}+\frac{\eta}{w}\right)\right\},
\label{eq:I}
\end{equation}
where $\omega=\sqrt{ab}$ and $\eta=\sqrt{a/b}$. For notational clarity, we will denote the parameterization given in \eqref{eq:I} by $\text{I}(\omega,\eta,\lambda)$.

\subsection{Skewed Matrix Variate Distributions}
The work of \cite{gallaugher17a,gallaugher17b} presents a total of four skewed matrix variate distributions, the matrix variate skew-$t$, generalized hyperbolic, variance-gamma and NIG distributions. Each of these distributions is derived from a matrix variate normal variance-mean mixture. In this representation, the random matrix $\fX$ has the representation
\begin{equation}
\fX=\matm+W\vecA+\sqrt{W}\fV,
\label{eq:mvmix}
\end{equation}
where $\matm$ and $\vecA$ are $n\times p$ matrices representing the location and skewness respectively, $\fV \sim \mathcal{N}_{n \times p}\left(\bf{0}, \matsig , \matPsi \right)$, and $W>0$ is a random variable with density $h(w|\btheta)$. 

In \cite{gallaugher17a}, the matrix variate skew-$t$ distribution, with $\nu$ degrees of freedom, is shown to arise as a special case of \eqref{eq:mvmix} with $W^{\text{ST}}\sim \text{IGamma}(\nu/2,\nu/2)$, where $\text{IGamma}(\cdot)$ denotes the inverse Gamma distribution with density 
$$
f(y~|~a,b)=\frac{b^a}{\Gamma(a)}y^{-a-1}\exp\left\{-\frac{b}{y}\right\}.
$$
The resulting density of $\fX$ is 
\begin{align*}
f_{\text{MVST}}(\vecX~|~\bvtheta)=&\frac{2\left(\frac{\nu}{2}\right)^{\frac{\nu}{2}}\exp\left\{\tr(\matsig^{-1}(\vecX-\matm)\matPsi^{-1}\vecA') \right\} }{(2\pi)^{\frac{np}{2}}| \matsig |^{\frac{p}{2}} |\matPsi |^{\frac{n}{2}}\Gamma(\frac{\nu}{2})}  \left(\frac{\delta(\vecX;\matm,\matsig,\matPsi)+\nu}{\rho(\vecA,\matsig,\matPsi)}\right)^{-\frac{\nu+np}{4}} \\ & \qquad\qquad\qquad\qquad\times
 K_{-\frac{\nu+np}{2}}\left(\sqrt{\left[\rho(\vecA,\matsig,\matPsi)\right]\left[\delta(\vecX;\matm,\matsig,\matPsi)+\nu\right]}\right),
\end{align*}
where 
$$
\delta(\vecX;\matm,\matsig,\matPsi)=\tr(\matsig^{-1}(\vecX-\matm)\matPsi^{-1}(\vecX-\matm)'),\quad  \rho(\vecA;\matsig,\matPsi)=\tr(\matsig^{-1}\vecA\matPsi^{-1}\vecA')
$$
and $\nu>0$.
For notational clarity, this distribution will be denoted by $\text{MVST}(\matm,\vecA,\matsig,\matPsi,\nu)$.

In \cite{gallaugher17b}, one of the distributions considered is a matrix variate generalized hyperbolic distribution. This again is the result of a special case of \eqref{eq:mvmix} with $W^{\text{GH}}\sim\text{I}(\omega,1,\lambda)$. This distribution will be denoted by $\text{MVGH}(\matm,\vecA,\matsig,\matPsi,\lambda,\omega)$, and the density is 
\begin{align*}
f_{\text{MVGH}}(\vecX|\bvtheta)=&\frac{\exp\left\{\tr(\matsig^{-1}(\vecX-\matm)\matPsi^{-1}\vecA') \right\}}{(2\pi)^{\frac{np}{2}}| \matsig |^{\frac{p}{2}} |\matPsi |^{\frac{n}{2}}K_{\lambda}(\omega)}  \left(\frac{\delta(\vecX;\matm,\matsig,\matPsi)+\omega}{\rho(\vecA,\matsig,\matPsi)+\omega}\right)^{\frac{\left(\lambda-\frac{np}{2}\right)}{2}} \\ & \times
 K_{\left(\lambda-{np}/{2}\right)}\left(\sqrt{\left[\rho(\vecA,\matsig,\matPsi)+\omega\right]\left[\delta(\vecX;\matm,\matsig,\matPsi)+\omega\right]}\right),
\end{align*}
where $\lambda\in \mathbb{R}$ and $\omega>0$.

The matrix variate variance-gamma distribution, also derived in \cite{gallaugher17b}, denoted $\text{MVVG}(\matm,\vecA,\matsig,\matPsi,\gamma)$ is a special case of the matrix variate normal variance-mean mixture \eqref{eq:mvmix} with $W^{\text{VG}}\sim\text{gamma}(\gamma,\gamma)$, where $\text{gamma}(\cdot)$ denotes the gamma distribution with density 
$$
f(y~|~a,b)=\frac{b^a}{\Gamma(a)}y^{a-1}\exp\left\{-by\right\}.
$$
The density of the random matrix $\fX$ with this distribution is
\begin{align*}
f_{\text{MVVG}}(\vecX|\bvtheta)=&\frac{2\gamma^{\gamma}\exp\left\{\tr(\matsig^{-1}(\vecX-\matm)\matPsi^{-1}\vecA') \right\}}{(2\pi)^{\frac{np}{2}}| \matsig |^{\frac{p}{2}} |\matPsi |^{\frac{n}{2}}\Gamma(\gamma)}  \left(\frac{\delta(\vecX;\matm,\matsig,\matPsi)}{\rho(\vecA,\matsig,\matPsi)+2\gamma}\right)^{\frac{\left(\gamma-{np}/{2}\right)}{2}} \\
&\times  K_{\left(\gamma-\frac{np}{2}\right)}\left(\sqrt{\left[\rho(\vecA,\matsig,\matPsi)+2\gamma\right]\left[\delta(\vecX;\matm,\matsig,\matPsi)\right]}\right),
\end{align*}
where $\gamma>0$.

Finally, the matrix variate NIG distribution arises when $W^{\text{NIG}}\sim\text{IG}(1,\tgamma)$, where $\text{IG}(\cdot)$ denotes the inverse-Gaussian distribution with density
$$
f(y~|~\delta,\gamma)=\frac{\delta}{\sqrt{2\pi}}\exp\{\delta\gamma\}y^{-\frac{3}{2}}\exp\left\{-\frac{1}{2}\left(\frac{\delta^2}{y}+\gamma^2y\right)\right\}.
$$
The density of $\fX$ is
\begin{align*}
f_{\text{MVNIG}}(\vecX|\bvtheta)&=\frac{2\exp\left\{\tr(\matsig^{-1}(\vecX-\matm)\matPsi^{-1}\vecA')+\tgamma\right\}
}{(2\pi)^{\frac{np+1}{2}}| \matsig |^{\frac{p}{2}} |\matPsi |^{\frac{n}{2}}}\left(\frac{\delta(\vecX;\matm,\matsig,\matPsi)+1}{\rho(\vecA,\matsig,\matPsi)+\tgamma^2}\right)^{-{\left(1+np\right)}/{4}}\\
&\times K_{-{(1+np)}/{2}}\left(\sqrt{\left[\rho(\vecA,\matsig,\matPsi)+\tgamma^2\right]\left[\delta(\vecX;\matm,\matsig,\matPsi)+1\right]}\right),
\end{align*}
where $\tgamma>0$. This distribution is denoted by $\text{MVNIG}(\matm,\vecA,\matsig,\matPsi,\tgamma)$. 

\subsection{Benefits Over Vectorization}
One alternative to matrix variate analysis for matrix variate data is to consider the vectorization of the data and perform multivariate techniques. However, the benefits of using matrix variate methods are twofold. The first being specifically for the case of multivariate longitudinal data. Performing the analysis using a matrix variate model has the benefit of simultaneously considering the temporal covariances (via $\matsig$) as well as the covariances for the variables (via $\matPsi$). Performing multivariate analysis on the vectorization of the data would not have this benefit without imposing some structure on the scale matrix.

The second benefit is the reduction in the number of parameters. If the matrix variate data is $n\times p$, vectorization would result in $np$ dimensional vectors, therefore resulting in $(n^2p^2+np)/2$ free scale parameters when using multivariate analysis. However, when using a matrix variate model, there are two lower dimensional matrices that comprise the scale parameters with a total of $(n^2+p^2+n+p)/2$ free scale parameters. Thus, for $n=p$, there is a reduction from quartic to quadratic complexity in $n$ and, for almost all values of $n$ and $p$, there will be a (often substantial) reduction in the number of free scale parameters.

\section{Methodology}
\subsection{Likelihoods}
In the mixture model context, $\fX$ is assumed to come from a population with $G$ subgroups each distributed according to the same one of the four skewed matrix variate distributions discussed previously. 
Now suppose $N$ $n\times p$ matrices $\vecX_1,\vecX_2,\ldots,\vecX_{N}$ are observed, then the observed-data likelihood is
\begin{align*}
L(\bvtheta)&=\prod_{i=1}^{N}\sum_{g=1}^G\pig f(\vecX_{i}~|~\matm_g,\vecA_g,\matsig_g,\matPsi_g,\btheta_g),
\end{align*}
where $\btheta_g$ are the parameters associated with the distribution of $W_{ig}$.
For the purposes of parameter estimation, we proceed as if the observed data is incomplete. In particular, we introduce the missing group membership indicators $z_{ig}$, where
$$z_{ig}=
\begin{cases}
1 & \text{if } \vecX_{i} \text{ is in group } g,\\
0 & \text{otherwise}. 
\end{cases}$$
In addition to the missing $z_{ig}$, we also have the latent variables $W_{ig}$ and we denote their densities by $h(w_{ig}~|~\btheta_g)$.

The complete-data log-likelihood, in its general form for any of the distributions already discussed, is then
\begin{equation}
\ell_c(\bvtheta)=\mathcal{L}_1+(\mathcal{L}_2+C_2)+(\mathcal{L}_3+C_3),
\label{eq:complik}
\end{equation}
where $C_2$ and $C_3$ are constant with respect to the parameters, $\mathcal{L}_1=\sum_{i=1}^N\sum_{g=1}^Gz_{ig}\pig$,\\ $\mathcal{L}_2=\sum_{i=1}^N\sum_{g=1}^Gz_{ig}h(w_{ig}~|~\btheta_g)-C_{2}$, and
\begin{align*}
\mathcal{L}_3=& \frac{1}{2}\sum_{i=1}^N\sum_{g=1}^Gz_{ig}\bigg[ \tr \left(\matsig_g^{-1}(\vecX_i-\matm_g)\matPsi_g^{-1}\vecA_g'\right)+\tr \left(\matsig_g^{-1}\vecA_g\matPsi_g^{-1}(\vecX_i-\matm_g)'\right)\\
&-\frac{1}{w_{ig}}\tr(\matsig_g^{-1}(\vecX_i-\matm_g)\matPsi_g^{-1}(\vecX_i-\matm_g)')-w_{ig}\tr(\matsig_g^{-1}\vecA_g\matPsi_g^{-1}\vecA_g')-p\log(|\matsig_g|)-n\log(|\matPsi_g|)\bigg].
\end{align*}
\subsection{Parameter Estimation}
Parameter estimation is performed by using an expectation-conditional maximization (ECM) algorithm \citep{meng93}. 

\noindent {\bf 1) Initialization}: Initialize the parameters $\matm_g,\vecA_g,\matsig_g,\matPsi_g$ and other parameters related to the distribution. Set $t=0$.

\noindent {\bf 2) E Step}: Update $\hat{z}_{ig}, a_{ig}, b_{ig}, c_{ig}$, where
\begin{equation*}\begin{split}
\hat{z}_{ig}^{(t+1)}&=\frac{\pig f(\vecX_i~|~\hat{\bvtheta}^{(t)}_g)}
{\sum_{h=1}^G\pi_h f(\vecX_i~|~\hat{\bvtheta}^{(t)}_h)},\qquad\qquad
a_{ig}^{(t+1)}=\mathbb{E}(W_{ig}|\vecX_i,z_{ig}=1,\hat{\bvtheta}_g^{(t)}),\\
b_{ig}^{(t+1)}&=\mathbb{E}\left(\frac{1}{W_{ig}}\left.\right|\vecX_i,z_{ig}=1,\hat{\bvtheta}_g^{(t)}\right),\qquad
c_{ig}^{(t+1)}=\mathbb{E}(\log(W_{ig})|\vecX_i,z_{ig}=1,\hat{\bvtheta}_g^{(t)}).\\
\end{split}\end{equation*}
%
%
Note that the specific updates will depend on the distribution. However, in each case, the conditional distribution of $W_{ig}$ given the observed data and group memberships is a generalized inverse Gaussian distribution. Specifically,
\begin{align*}
W_{ig}^{\text{ST}}~|~\vecX_i, z_{ig}=1&\sim \text{GIG}\left(\rho(\vecA_g,\matsig_g,\matPsi_g),\delta(\vecX;\matm_g,\matsig_g,\matPsi_g)+\nu_g,-(\nu_g+np)/2\right),\\
W_{ig}^{\text{GH}}~|~\vecX_i, z_{ig}=1&\sim \text{GIG}\left(\rho(\vecA_g,\matsig_g,\matPsi_g)+\omega_g,\delta(\vecX;\matm_g,\matsig_g,\matPsi_g)+\omega_g,\lambda_g-{np}/{2}\right),\\
W_{ig}^{\text{VG}}~|~\vecX_i, z_{ig}=1&\sim \text{GIG}\left(\rho(\vecA_g,\matsig_g,\matPsi_g)+2\gamma_g,\delta(\vecX;\matm_g,\matsig_g,\matPsi_g),\gamma_g-{np}/{2}\right),\\
W_{ig}^{\text{NIG}}~|~\vecX_i, z_{ig}=1&\sim \text{GIG}\left(\rho(\vecA_g,\matsig_g,\matPsi_g)+\tgamma_g^2,\delta(\vecX;\matm_g,\matsig_g,\matPsi_g)+1,-{(1+np)}/{2}\right).
\end{align*}
Therefore, the exact updates are obtained by using the expectations given in \eqref{eq:ai}--\eqref{eq:ci} for appropriate values of $\lambda, a$, and $b$. 

\noindent {\bf 3) First CM Step}: Update the parameters $\pig,\matm_g,\vecA_g$. 
\begin{align*}
\hat{\pi}_g^{(t+1)}&=\frac{N_g}{N}, \qquad\qquad 
\hat{\matm}_g^{(t+1)}=\frac{\sum_{i=1}^N\hat{z}_{ig}^{(t+1)}\vecX_i\left(\overline{a}_g^{(t+1)}b^{(t+1)}_{ig}-1\right)}{\sum_{i=1}^N\hat{z}_{ig}^{(t+1)}\overline{a}_g^{(t+1)}b_{ig}^{(t+1)}-N_g}, \\ 
\hat{\vecA}^{(t+1)}&=\frac{\sum_{i=1}^N\hat{z}_{ig}^{(t+1)}\vecX_i\left(\overline{b}_g^{(t+1)}-b^{(t+1)}_{ig}\right)}{\sum_{i=1}^N\hat{z}_{ig}^{(t+1)}\overline{a}_g^{(t+1)}b_{ig}^{(t+1)}-N_g}, 
\end{align*}
where 
$$
N_g=\sum_{i=1}^N\hat{z}_{ig}^{(t+1)},\qquad\overline{a}_g^{(t+1)}=\frac{\sum_{i=1}^N\hat{z}_{ig}^{(t+1)}a_{ig}^{(t+1)}}{N_g},\qquad \overline{b}_g^{(t+1)}=\frac{\sum_{i=1}^N\hat{z}_{ig}^{(t+1)}b_{ig}^{(t+1)}}{N_g}.
$$

\noindent {\bf 4) Second CM Step}: Update $\matsig_g$
\begin{equation}
\begin{split}
\hat{\matsig}_g^{(t+1)}&=\frac{1}{N_gp}\left[\sum_{i=1}^N\hat{z}_{ig}^{(t+1)}\left(b^{(t+1)}_{ig}\left(\vecX_i-\hat{\matm}_g^{(t+1)}\right)\left.\hat{\matPsi}_g^{(t)}\right.^{{-1}}\left(\vecX_i-\hat{\matm}_g^{(t+1)}\right)'\right.\right.\\&\left.\left.-\left.\hat{\vecA}_g^{(t+1)}\right.\left.\hat{\matPsi}_g^{(t)}\right.^{{-1}}\left(\vecX_i-\hat{\matm}_g^{(t+1)}\right)'
-\left(\vecX_i-\hat{\matm}_g^{(t+1)}\right)\left.\hat{\matPsi}_g^{(t)}\right.^{{-1}}\left.\hat{\vecA}_g^{(t+1)}\right.'\right.\right.\\&+\left.\left.a_{ig}^{(t+1)}\left.\hat{\vecA}_g^{(t+1)}\right.\left.\hat{\matPsi}_g^{(t)}\right.^{{-1}}\left.\hat{\vecA}_g^{(t+1)}\right.'\right) \right].
\end{split}
\end{equation}

\noindent {\bf 5) Third CM Step}: Update $\matPsi_g$
\begin{equation}
\begin{split}
\hat{\matPsi}_g^{(t+1)}&=\frac{1}{N_gn}\left[\sum_{i=1}^N\hat{z}_{ig}^{(t+1)}\left(b^{(t+1)}_{ig}\left(\vecX_i-\hat{\matm}_g^{(t+1)}\right)'\left.\hat{\matsig}_g^{(t+1)}\right.^{{-1}}\left(\vecX_i-\hat{\matm}_g^{(t+1)}\right)\right.\right.\\&\left.\left.-\left.\hat{\vecA}_g^{(t+1)}\right.'\left.\hat{\matsig}_g^{(t+1)}\right.^{{-1}}\left(\vecX_i-\hat{\matm}_g^{(t+1)}\right)
-\left(\vecX_i-\hat{\matm}_g^{(t+1)}\right)'\left.\hat{\matsig}_g^{(t+1)}\right.^{{-1}}\left.\hat{\vecA}_g^{(t+1)}\right.\right.\right.\\&+\left.\left.a_{ig}^{(t+1)}\left.\hat{\vecA}_g^{(t+1)}\right.'\left.\hat{\matsig}_g^{(t+1)}\right.^{{-1}}\left.\hat{\vecA}_g^{(t+1)}\right. \right)\right].
\end{split}
\end{equation}

\noindent {\bf 6) Other CM Steps}:
The additional parameters introduced by the distribution of $W_{ig}$ are now updated. These updates will vary according the distribution and the particulars for the MVST, MVGH, MVVG and MVNIG distributions are given below.

\noindent {\bf 7) Check Convergence}: If not converged, set $t=t+1$ and return to step 2).

\subsubsection*{Matrix Variate Skew-$t$ Distribution} 

In the case of the matrix variate skew-$t$ distribution, the degrees of freedom $\nu_g$ need to be updated. This update cannot be obtained in closed form, and thus needs to be performed numerically. We have
$$
\mathcal{L}_{2}^{\text{MVST}}=\sum_{i=1}^N\sum_{g=1}^Gz_{ig}\left[\frac{\nu_g}{2}\log\left(\frac{\nu_g}{2}\right)-\log\left(\Gamma\left(\frac{\nu_g}{2}\right)\right)-\frac{\nu_g}{2}\left(\log(w_{ig})+\frac{1}{w_{ig}}\right)\right].
$$
Therefore, the update $\nu_g^{(t+1)}$ is obtained by solving \eqref{eq:nugup} for $\nu_g$, i.e.,
\begin{equation}
\log\left(\frac{\nu_g}{2}\right)+1-\varphi\left(\frac{\nu_g}{2}\right)-\frac{1}{N_g}\sum_{i=1}^N\hat{z}_{ig}^{(t+1)}(b^{(t+1)}_{ig}+c^{(t+1)}_{ig})=0,
\label{eq:nugup}
\end{equation}
where $\varphi(\cdot)$ denotes the digamma function.

\subsubsection*{Matrix Variate Generalized Hyperbolic Distribution}

In the case of the matrix variate generalized hyperbolic distribution, updates for $\lambda_g$ and $\omega_g$ are needed. In this case,
\begin{equation}
\mathcal{L}_2^{\text{MVGH}}=\sum_{i=1}^N\sum_{g=1}^Gz_{ig}\left[\log(K_{\lambda_g}(\omega_g))-\lambda_g\log w_{ig}-\frac{1}{2}\omega_g\left(w_{ig}+\frac{1}{w_{ig}}\right)\right].
\label{eq:L1GH}
\end{equation}
The updates for $\lambda_g$ and $\omega_g$ cannot be obtained in closed form. However, \cite{browne15} discuss numerical methods for these updates and, because the portion of the likelihood function that include these parameters is the same as in the multivariate case, the updates described in \cite{browne15} can be used directly here.

The updates for $\lambda_g$ and $\omega_g$ rely on the log-convexity of $K_{s}(t)$ in both $s$ and $t$ \citep{baricz10} and maximizing \eqref{eq:L1GH} via conditional maximization.
The resulting updates are
\begin{align}
\hat{\lambda}_g^{(t+1)}&=\bar{c}_g^{(t+1)}\hat{\lambda}_g^{(t)}\left[\left.\frac{\partial}{\partial s}\log(K_{s}(\hat{\omega}_g^{(t)}))\right|_{s=\hat{\lambda}_g^{(t)}}\right]^{-1}, \label{eq:lamup}\\
\hat{\omega}_g^{(t+1)}&=\hat{\omega}_g^{(t)}-\left[\left.\frac{\partial}{\partial s}q(\hat{\lambda}_g^{(t+1)},s)\right|_{s=\hat{\omega}_g^{(t)}}\right]\left[\left.\frac{\partial^2}{\partial s^2}q(\hat{\lambda}_g^{(t+1)},s)\right|_{s=\hat{\omega}_g^{(t)}}\right]^{-1}, \label{eq:omup}
\end{align}
where the derivative in \eqref{eq:lamup} is calculated numerically, 
$$
q(\lambda_g,\omega_g)=\sum_{i=1}^Nz_{ig}\left[\log(K_{\lambda_g}(\omega_g))-\lambda_g\log w_{ig}-\frac{1}{2}\omega_g\left(w_{ig}+\frac{1}{w_{ig}}\right)\right]
$$
 and $\bar{c}_g^{(t+1)}=({1}/{N_g})\sum_{i=1}^N\hat{z}^{(t+1)}_{ig}c_{ig}^{(t+1)}$. The partials in \eqref{eq:omup} are described in \cite{browne15} and can be written as
$$
\frac{\partial}{\partial \omega_g}q(\lambda_g,\omega_g)=\frac{1}{2}[R_{\lambda_g}(\omega_g)+R_{-\lambda_g}(\omega_g)-(\bar{a}_g^{(t+1)}+\bar{b}_g^{(t+1)})],
$$
and 
$$
\frac{\partial^2}{\partial \omega_g^2}q(\lambda_g,\omega_g)=\frac{1}{2}\left[R_{\lambda_g}(\omega_g)^2-\frac{1+2\lambda_g}{\omega_g}R_{\lambda_g}(\omega_g)-1+R_{-\lambda_g}(\omega_g)^2-\frac{1-2\lambda_g}{\omega_g}R_{-\lambda_g}(\omega_g)-1\right],
$$
where $R_{\lambda_g}(\omega_g)=K_{\lambda_g+1}(\omega_g)/K_{\lambda_g}(\omega_g)$.

\subsubsection*{Matrix Variate Variance-Gamma Distribution}

In the case of the matrix variate variance-gamma, 
$$
\mathcal{L}_2^{\text{MVVG}}=\sum_{i=1}^N\sum_{g=1}^Gz_{ig}\left[\gamma_g\log\gamma_g-\log\Gamma(\gamma_g)+\gamma_g\left(\log w_{ig}-w_{ig}\right)\right].
$$
The update for $\gamma_g$, as in the skew-t case, cannot be obtained in closed form. Instead, the update $\gamma_g^{(t+1)}$ is obtained by solving \eqref{eq:gammup} for $\gamma_g$, where
\begin{equation}
\log\gamma_g+1-\varphi(\gamma_g)+\bar{c}_g^{(t+1)}-\bar{a}_g^{(t+1)}=0.
\label{eq:gammup}
\end{equation}

\subsubsection*{Matrix Variate NIG Distribution}
 
In this case, $\tgamma_g$ needs to be updated. Note that
$$
\mathcal{L}_2^{\text{MVNIG}}=\sum_{i=1}^N\sum_{g=1}^Gz_{ig}\tgamma_g-\frac{\tgamma_g^2}{2}z_{ig}w_{ig}
$$
and, therefore, the closed form updates for $\tgamma_g$ are
$$
\tgamma_{g}^{(t+1)}=\frac{N_{g}}{\overline{a}^{(t+1)}_g}.
$$

\subsection{Numerical Considerations and Convergence Criterion}
The main numerical problem encountered is the calculation of the Bessel function $K_{\lambda}(x)$ and the calculation of its partial derivative with respect to $\lambda$. When $x$ becomes large relative to $\lambda$, the Bessel function is rounded to zero. One solution is to consider the evaluation of $\exp\{x\}K_{\lambda}(x)$ and then make adjustments to subsequent calculations. In most of the simulations (Section~4), this helped with the evaluation of the densities and the numerical derivatives. However, for the real data application (Section~5), the issue is that $|\lambda|$ becomes too large due to the dimension and the Bessel function tends to infinity. This is an indication that dimension reduction techniques will need to be considered in the future (see Section~\ref{sec:disc}).

There are several options for determining convergence of this ECM algorithm. The criterion used in the simulations in Section~4 is based on the Aitken acceleration \cite{aitken26}. The Aitken acceleration at iteration $t$ is
$$
a^{(t)} = \frac{l^{(t+1)}-l^{(t)}}{l^{(t)}-l^{(t-1)}},
$$
where $l^{(t)}$ is the (observed) log-likelihood at iteration $t$.  We then define 
$$l_{\infty}^{(t+1)} = l^{(t)} + \frac{1}{1-a^{(t)}}(l^{(t+1)}-l^{(t)})$$ \citep[see][]{bohning94, lindsay95}. The quantity $l_{\infty}$ is an asymptotic estimate (i.e., an estimate of the value after many iterations) of the log-likelihood at iteration $t+1$. As in \cite{mcnicholas10a}, we stop our EM algorithms when \begin{equation}\label{eqn:mcn}l_{\infty}^{(k+1)}-l^{(k)} < \epsilon,\end{equation} provided this difference is positive. The main benefit of this criterion when compared to lack of progress, is that the likelihood can sometimes ``plateau" before increasing again. Accordingly, if lack of progress is used, the algorithm may terminate prematurely \citep[see][]{mcnicholas10a}. However, the criterion in \eqref{eqn:mcn} helps overcome this problem by considering the likelihood after very many iterations, i.e., $l_{\infty}$.

\subsection{A Note on Identifiability}
It is important to note that, for each of the distributions discussed herein, the estimates for $\matsig_g$ and $\matPsi_g$ are only unique up to a strictly positive constant. Therefore, to eliminate the identifiability issue, a constraint needs to be imposed on $\matsig_g$ or $\matPsi_g$. \cite{Anderlucci15}, suggest taking the trace of $\matPsi_g$ to be equal to $p$; however, it is much simpler to set the first diagonal element of $\matsig_g$ to be 1 and this is the constraint we use in the analyses herein. 

Discussion of identifiability would not be complete without mention of the label switching problem. This well-known problem is due to the invariance of the mixture model to relabelling of the components \citep{render84,stephens00}. While the label switching problem is a real issue in the Bayesian paradigm \cite[see][for some discussion]{stephens00,celeux00}, it is of no practical concern for the work carried out herein. However, it is a theoretical identifiability issue and we note that it be resolved by specifying some ordering on the model parameters, e.g., simply requiring that $\pi_1>\pi_2>\cdots>\pi_G$ often works and ordering on other parameters can be imposed as needed.

\subsection{Number of Components and Performance Evaluation}
In a general clustering scenario, the number of components (groups) $G$ are not known {\it a priori}. It is, therefore, necessary to select an adequate number of components. There are two methods that are quite common in the literature. The first is the Bayesian information criterion  \citep[BIC;][]{schwarz78}, which is defined as
$$
\text{BIC}=2{\ell}(\hat{\bvtheta})-p\log N,
$$
where ${\ell}(\hat{\bvtheta})$ is the maximized log-likelihood, $N$ is the number of observations, and $p$ is the number of free parameters. 
Another criterion common in the literature is the integrated completed likelihood \citep[ICL;][]{biernacki00}. The ICL can be approximated as
$$
\text{ICL}\approx\text{BIC}+2\sum_{i=1}^{N}\sum_{g=1}^G\text{MAP}(\hat{z}_{ig})\log\hat{z}_{ig},
$$
where 
$$\text{MAP}(\hat{z}_{ig})=
\begin{cases}
1 & \text{if arg max}_{h=1,\ldots,G}\left\{\hat{z}_{ih}\right\}=g,\\
0 & \text{otherwise}.
\end{cases}$$
The ICL can be viewed as penalized version of the BIC, where the penalty is for uncertainty in the component membership.

To evaluate clustering performance, we consider the adjusted Rand index \citep[ARI;][]{hubert85}. The ARI compares two different partitions, i.e., two different classifications in our applications. When the predicted classification is compared the actual classification, an ARI of 1 corresponds to perfect classification, whereas a value of 0 indicates the predicted classification is no better than randomly assigning the labels. Detailed review and discussion of the ARI is provided by \cite{steinley04}.

\subsection{Semi-Supervised Classification}
In addition to clustering (unsupervised classification), the matrix variate mixture models introduced here can also be applied for semi-supervised classification. Suppose that $N$ matrices are observed but that we know the labels for $K$ of the $N$ matrices; specifically, suppose that $K$ of the $N$ matrices come from one of $G$ classes. By analogy with \cite{mcnicholas10c}, and without loss of generality, order these matrices so it is the first $K$ that have known labels: $\vecX_1,\ldots,\vecX_K,\vecX_{K+1},\ldots\vecX_{N}$. Now, we know the values of $z_{ig}$ for $i=1,\ldots,K$ and the observed-data likelihood is
\begin{align*}
L(\bvtheta)&=\prod_{i=1}^{K}\prod_{g=1}^G\left[\pig f(\vecX_{i}~|~\matm_g,\vecA_g,\matsig_g,\matPsi_g,\btheta_g)\right]^{z_{ig}}
\times\prod_{j=K+1}^{N}\sum_{h=1}^H\pig f(\vecX_{j}~|~\matm_h,\vecA_h,\matsig_h,\matPsi_h,\btheta_h),
\end{align*}
where $\btheta_g$ are the parameters associated with the distribution of $W_{ig}$. In general, $H\geq G$; however, for the analyses herein, we make the common assumption that $H=G$. Parameter estimation, identifiability, etc., follow in an analogous fashion to the clustering case already described herein. Further details on semi-supervised classification in the mixture model setting are given in \cite{mclachlan00} and \cite{mcnicholas16a}.

\section{Simulations}
\subsection{Overview}
Two simulations are performed, where the first simulation has two groups and the second has three. The chosen parameters have no intrinsic meaning; however, they can be viewed as representations of multivariate longitudinal data and the parameters introduced by the distribution of $W_{ig}$ are meant to illustrate the flexibility in concentration. Simulation~1 considers $3\times 4$ data, Simulation~2 illustrates $4\times 3$ data. In the first simulation, $\matsig_g$ and $\matPsi_g$ are set to
$$
\begin{array}{ll}
\matsig_1=\left(
\begin{array}{ccc}
1&0.5&0.1\\
0.5 & 1 &0.5\\
0.1& 0.5& 1\\
\end{array}
\right),
&
\matsig_2=\left(
\begin{array}{ccc}
1&0.1&0.1\\
0.1 & 1 &0.1\\
0.1& 0.1& 1\\
\end{array}
\right),
\\
\\[-12pt]
\matPsi_1=\left(
\begin{array}{cccc}
1& 0.5&0.5 &0.5\\
0.5 & 1 & 0&0\\
0.5 &0 &1&0\\
0.5&0&0&1\\
\end{array}
\right),&
\matPsi_2=\left(
\begin{array}{cccc}
1& 0&0 &0\\
0 & 1 & 0.5&0.5\\
0 &0.5 &1&0.2\\
0&0.5&0.2&1\\
\end{array}
\right).
\end{array}
$$

For notational purposes, let $\tilde{\matsig}_g$ and $\tilde{\matPsi}_g$ be the scale matrices used in Simulation 2. We set $\tilde{\matsig}_1=\matPsi_1$, $\tilde{\matsig}_2=\tilde{\matsig}_3=\matPsi_2$ and $\tilde{\matPsi}_1=\tilde{\matPsi}_3=\matsig_1$ and $\tilde{\matPsi}_2=\matsig_2$.
For each distribution, the models are fitted for $G\in\{1,2,3,4\}$ and the BIC is used to choose the number of groups.

\subsection{Simulation 1}
In Simulation 1, for all four distributions, we take the location and skewness matrices to be:
$$
\begin{array}{ll}
\matm_1=\left(
\begin{array}{rrrr}
1   & 0&    0&   -1\\
0   & 1   &-1 &   0\\
-1   & 0    &2 &  -1
\end{array}
\right),&
\matm_2=\left(
\begin{array}{llll}
3 & 4 & 2 & 4 \\ 
   4 & 3 & 3 & 3 \\ 
   3 & 4 & 2 & 4 \\ 
\end{array}
\right),\\
\\[-12pt]
\vecA_1=\left(
\begin{array}{llll}
1 & -1 & 0 & 1 \\ 
   1 & -1 & 0 & 1 \\ 
  1 & -1 & 0 & 1 \\ 
\end{array}
\right),&
\vecA_2=\left(
\begin{array}{llll}
1 & 1 & 1 & -1 \\ 
   1 & 1 & 0.5 & -1 \\ 
  1 & 1 & 0 & -1 \\ 
\end{array}
\right).\\
\end{array}
$$
For the additional parameters, we took $\nu_1=4,\nu_2=20$ for the skew-$t$ distribution, $\lambda_1=\lambda_2=2$ and $\omega_1=4$, $\omega_2=2$ for the generalized hyperbolic distribution, $\gamma_1=7$, $\gamma_2=14$ for the variance-gamma distribution, and $\tgamma_1=1/2,\tgamma_2=2$ for the NIG distribution. Figure \ref{fig:Sim1} in Appendix A shows a typical dataset for each distribution. For visualization, we look at the marginal columns which we label V1, V2, V3 and V4. We see that for all of the columns, except column 4, there is a clear separation between the two groups. We also note that for the skew-$t$ distribution, there was a severe outlier in group 2 (due to the small degrees of freedom) that we do not show for better visualization. The orange dotted line is the marginal location parameter for the first group, and the yellow dotted line is the marginal location for the second group. 

Table \ref{tab:ResSim1} displays the number of groups (components) chosen and the average ARI values with the associated standard deviations. The ICL results are identical, and thus are not shown here. We see that the correct number of groups is chosen, with perfect classification, for all 30 of the datasets when using the MVST, MVVG, and MVNIG mixtures. However, this is not the case with MVGH mixture, which underperforms when compared to the other three. However, the eight datasets for which the incorrect number of components is chosen correspond to datasets for which the two-component MVGH solution did not converge and, in a real application, alternative starting values would be pursued until convergence is achieved for the $G=2$ component case.  
\begin{table}[!htb]
\centering
\caption{The number of groups chosen by the BIC and the average ARI values, with standard deviations in parentheses, for Simulation 1. Note that the MVGH mixture did not converge for eight of the 30 runs with $G=2$.}
\begin{tabular}{lccccr}
\hline
 &$G=1$&$G=2$ &$G=3$ &$G=4$ & $\overline{\text{ARI}}$ (std.\ dev.)\\
\hline
MVST & 0 & 30 & 0 & 0 & 1.00 (0.00)\\
MVGH & 4 & 22 & 1 &3&0.85 (0.34)\\
MVVG & 0 & 30 & 0&0 & 1.00 (0.00)\\
MVNIG & 0 & 30 & 0&0&1.00 (0.00)\\
\hline
\end{tabular}
\label{tab:ResSim1}
\end{table}

In Table \ref{tab:Time_Sim1}, we show the average amount of time per dataset to run the algorithm for $G=1, 2, 3, 4$. We note that these simulations were performed in parallel.
\begin{table}[!hbt]
\centering
\caption{Average runtimes for Simulation 1.}
\begin{tabular}{lr}
\hline
Distribution & Average Time (s)\\
\hline
MVST & 237.33 \\
MVGH & 625.90\\
MVVG & 82.77\\
MVNIG & 349.47\\
\hline
\end{tabular}
\label{tab:Time_Sim1}
\end{table}

\subsection{Simulation 2}
In Simulation 2, a three group mixture was considered with 200 observations per group and the following location and skewness parameters.
$$
\begin{array}{lll}
\matm_1=\left(
\begin{array}{rrr}
 1 & -1 & 0 \\ 
   0 & 0 & -1 \\ 
   0 & 1 & 0 \\ 
 -1 & 0 & -1 \\  
\end{array}
\right),&
\matm_2=\left(
\begin{array}{cccc}
-1 & 1 & 0 \\ 
  0 & 0 & 1 \\ 
  0 & -1 & 0 \\ 
  1 & 0 & 1 \\ 
\end{array}
\right),&
\matm_3=\left(
\begin{array}{cccc}
1 & 1 & 2 \\ 
  1 & 2 & 0 \\ 
  0 & 1 & 1 \\ 
  0 & 1 & 0 \\ 
\end{array}
\right),
\\
\\[-12pt]
\vecA_1=\left(
\begin{array}{lll}
1 & -1 & -1 \\ 
  1 & -0.5 & -1 \\ 
  1 & 0 & -1 \\ 
  1 & 0 & -1 \\ 
\end{array}
\right),&
\vecA_2=\vecA_3=\left(
\begin{array}{lll}
1 & 1 & -1 \\ 
  1 & 0.5 & 0.5 \\ 
  1 & 0 & 0 \\ 
  1 & 0 & 0 \\ 
\end{array}
\right).\\
\end{array}
$$
The other parameters we set to $\nu_1=4$, $\nu_2=8$, $\nu_3=20$ for the MVST, $\lambda_1=4$, $\lambda_2=0$, $\lambda_3=-2$ and $\omega_1=4$, $\omega_2=\omega_3=2$ for the MVGH, $\gamma_1=7$, $\gamma_2=9$, $\gamma_3=14$ for the MVVG and $\tgamma_1=1/2$, $\tgamma_2=1$, $\tgamma_3=2$ for the MVNIG.

Again, the marginal distributions of a typical dataset is shown in Figure~\ref{fig:Sim2} in Appendix~A. The dotted lines again represent the marginal locations, with orange for the first group, yellow for the second, and purple for the third. In Table~\ref{tab:ResSim2}, the number of groups chosen by the BIC as well as the average ARI values, and associated standard deviations, are presented. As before, the MVST, MVVG and MVNIG mixtures outperform the MVGH mixture and, once again, this is due to convergence issues. The issue with convergence for the MVGH mixture with both simulations is possibly due to the update for, or impact of, the index parameters $\lambda_1,\ldots,\lambda_G$. 
\begin{table}[!htb]
\centering
\caption{The number of groups chosen by the BIC and the average ARI values, with standard deviations in parentheses, for Simulation 2. Note that the MVGH mixture did not converge for 22 of the 30 runs with $G=2$.}
\begin{tabular}{lccccr}
\hline
 &$G=1$&$G=2$ &$G=3$ &$G=4$ & $\overline{\text{ARI}}$ (std.\ dev.)\\
\hline
MVST & 0 & 0 & 30 & 0 & 0.97 (0.010)\\
MVGH & 10 & 8 & 8 &4&0.52 (0.41)\\
MVVG & 0 & 0 & 30 &0 & 0.98 (0.0077)\\
MVNIG & 0 & 0 & 30 &0 &0.99 (0.0056)\\
\hline
\end{tabular}
\label{tab:ResSim2}
\end{table}

Table \ref{tab:Time_Sim2} shows the average runtime per dataset for Simulation~2. Notice that for the MVGH, MVVG and MVNIG mixtures, each dataset took longer on average, with the MVGH mixture having the longest runtime as well as the largest increase in runtime over Simulation~1. This is to be expected because there is an increase in the number of groups and observations; however, for the MVVG and MVNIG mixtures, the time differences between Simulations~1 and~2 are less notable. The MVST mixture actually took less time on average; however, this is because a few datasets for Simulation 1 ran to the maximum number of iterations (without converging) for the $G=4$ group mixture thus increasing the runtime. 
\begin{table}[!htb]
\centering
\caption{Average runtimes for Simulation 2.}
\begin{tabular}{lr}
\hline
Distribution & Average Time (s)\\
\hline
MVST & 233.67 \\
MVGH & 2542.50\\
MVVG & 171.90\\
MVNIG & 581.63\\
\hline
\end{tabular}
\label{tab:Time_Sim2}
\end{table}

\section{Image Recognition Example}
We now apply the matrix variate mixture models introduced herein to image recognition with the MNIST handwriting dataset \citep{MNIST}. The original dataset consists of 60,000 training images of handwritten digits 0 to 9, which can be represented as $28 \times 28$ pixel matrices with greyscale intensities ranging from 0 to 255. However, these dimensions resulted in an infinite calculation for the Bessel function and its derivative with respect to $\lambda$. Moreover, because two unstructured $28\times 28$ dimensional covariance matrices would need to be estimated, model fitting would be infeasible. We stress that this alone is an indication that dimension reduction techniques will need to be developed in the future. However, the main goal of this application is to demonstrate the discussed methods outside of the theoretical confines of the simulations. Therefore, we resized the original image to a $10\times 10$ pixel matrix using the {\it resize} function in the {\tt EBImage} package \citep{EBImage} for the {\sf R} software \citep{R16}. 
However, there are problems with sparsity. Specifically, the outside columns and rows all contain values of 0 because they are outside of the main writing space. Accordingly, there is no variation in these outer columns and rows, therefore resulting in exactly singular $\matsig_g$ and $\matPsi_g$ updates. To solve this problem, we replace a value of 0 with a value between 0 and 2 with increments of 0.1 and added 50 to the non-zero values to make sure the noise did not interfere with the true signal. 

Each of the matrix variate mixtures introduced herein is applied within the semi-supervised classification paradigm (Section~3.6). A total of 500 observations from digit~1 and 500 from digit~7 are sampled from the training set, and then 100 of each of these digits is  considered unlabelled, i.e., 80\% of the data are labelled. We performed the analysis on 30 different such sets. In Table~\ref{tab:App1vs2}, we show aggregate classification tables for the points considered unlabelled for each of the matrix variate mixtures. In Table \ref{tab:App1vs2ARI}, we show the average ARI values and the average misclassification rates for the unlabelled points. Note, that for some of the datasets, not all four mixtures converged; therefore, the total number of observations in the tables need not be the same for all four distributions. Looking at the classification tables, it is clear that all of these matrix variate mixtures overall misclassify digit~1 as digit~7 more often than digit~7 as digit~1. From both the ARI and MCR results, the MVVG mixture slightly outperforms the other three mixtures. It is interesting to note that the MVGH mixture did not experience the same convergence issues as seen with the simulations. This is almost certainly because 80\% of the data points have known labels here whereas, in the simulations, we used the clustering (unsupervised classification) scenario and so all of the labels are unknown.
\begin{table}[!htb]
\centering
\caption{Cross-tabulations of true (1,7) versus predicted (P1, P7) classifications for the points considered unlabelled in the MNIST data, for each of the matrix variate mixtures introduced herein, aggregated over all runs (for which convergence was attained).}
\begin{tabular}{rcc|cc|cc|cc}
\hline
&\multicolumn{2}{c|}{MVST}&\multicolumn{2}{|c|}{MVGH}&\multicolumn{2}{|c|}{MVVG}
&\multicolumn{2}{|c}{MVNIG}\\
\hline
& P1&P7&P1&P7&P1&P7&P1&P7\\
\hline
1 &  2797 & 203 & 2813&187 & 2859 & 141&2798 & 202 \\ 
   7 &127 & 2873 & 125&2875& 122 & 2878&127 & 2873\\
\hline
\end{tabular}
\label{tab:App1vs2}
\end{table} 
\begin{table}[!ht]
\centering
\caption{Average ARI values and misclassification rates (MCR), with associated standard deviations in parentheses, for each matrix variate mixture approach for the points considered unlabelled for the MNIST data, aggregated over all runs (for which convergence was attained).}
\begin{tabular}{lrr}
\hline
 & $\overline{\text{ARI}}$ (std.\ dev.) &$\overline{\text{MCR}}$ (std.\ dev.)\\ 
\hline
MVST & 0.79 (0.051) & 0.055 (0.014)\\
MVGH & 0.80 (0.056) & 0.052 (0.016)\\
MVVG & 0.83 (0.043) & 0.044 (0.012)\\
MVNIG & 0.79 (0.051) & 0.055 (0.014)\\
\hline
\end{tabular}
\label{tab:App1vs2ARI}
\end{table}

\section{Discussion}\label{sec:disc}
Four matrix variate mixture distributions, with component densities that parameterize skewness, have been used for model-based clustering --- and its semi-supervised analogue --- of three-way data. Specifically, we considered MVST, MVGH, MVVG, and MVNIG mixtures, respectively, and an ECM algorithm was used for parameter estimation in each case. Simulated and real data were used for illustration. In the first simulation, there was good separation between the two groups and, in the second, we increased the number of groups, decreased the separation between the groups, and obtained similar results to the first. In both simulations, the MVGH mixture often underperformed when compared to the other three mixtures due to convergence issues. This could be resolved, for example, by restricting the index parameter $\lambda$; however, doing this would essentially eliminate the additional flexibility enjoyed by the MVGH mixture. In the real data application, the MVVG mixture outperformed the other three mixtures in terms of both average ARI and average misclassification rate, and the MVVG mixture consistently ran faster than the other three mixtures.

The next step in this work is to introduce parsimony into the matrix variate mixture models introduced herein. A very simple way to introduce parsimony is to take the eigenvalue decomposition of the scale matrices to form a family of parsimonious mixture models, along similar lines to \citep{celeux95}. Another important area, though slightly more difficult, is dimension reduction. Recall, in the MNIST data application, that the original data had to be resized due to problems evaluating the Bessel function as well as feasibility issues. One possible solution is to consider a matrix variate analogue of the mixture of factor analyzers model \citep{ghahramani97} and this will be a topic of future work. Another possibility for future work is the application of these models to multivariate longitudinal data \citep[e.g., as in][]{Anderlucci15}, in which case it would be important to impose a structure on $\matsig_g$. Finally, the unsupervised and semi-supervised classification paradigms have been investigated herein but some future work will focus on applying these matrix variate mixtures within the fractionally-supervised classification framework \citep{vrbik15,gallaugher17c}.

{\small
\section*{Acknowledgements}
{The authors gratefully acknowledge the support of a Vanier Canada Graduate Scholarship (Gallaugher) and the Canada Research Chairs program (McNicholas).}

}

\appendix
\section{Figures}
\begin{figure}[!htb]
\centering
\includegraphics[width=0.85\textwidth]{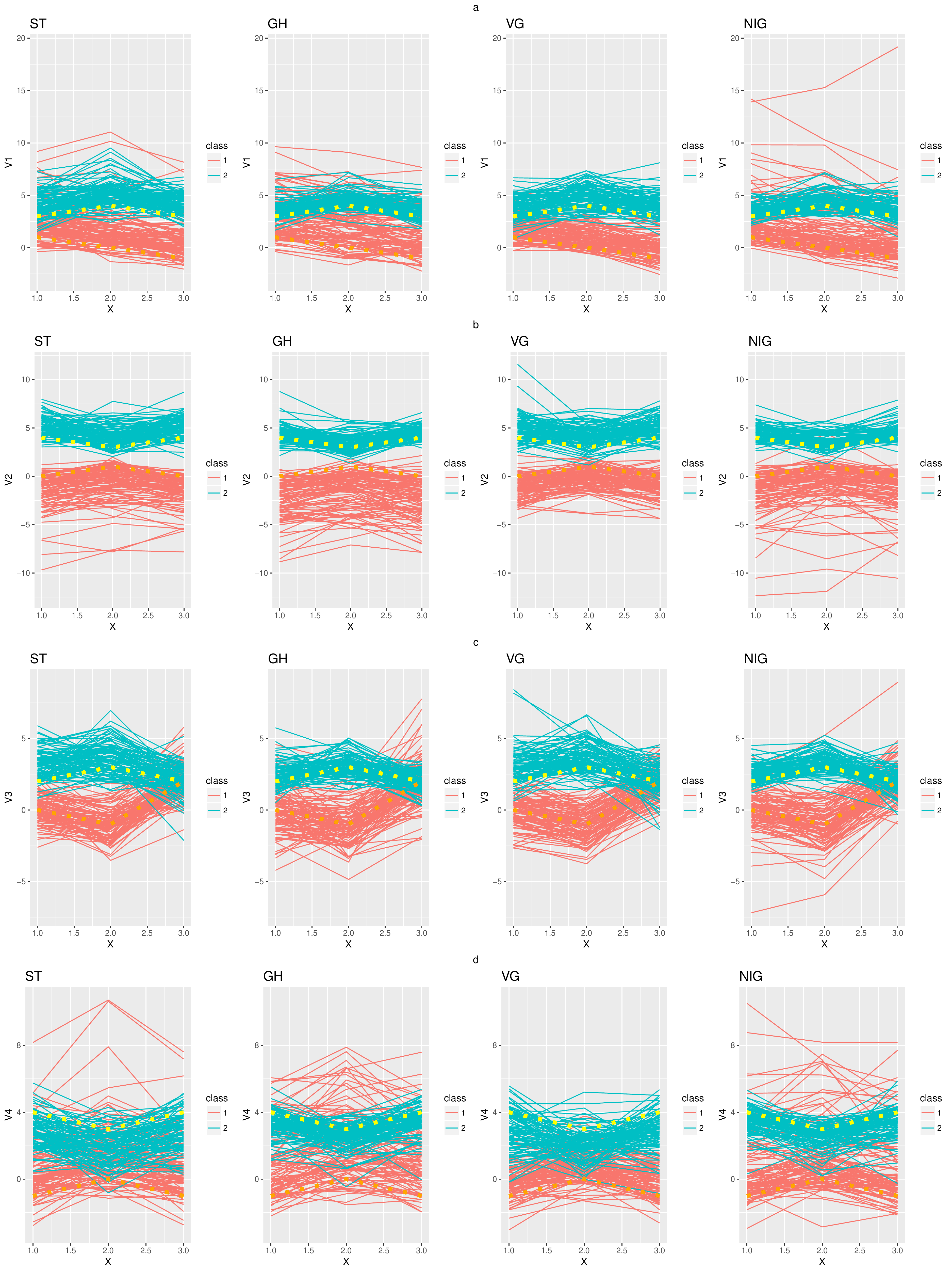}
\caption{Marginal data for the columns for each of the four distributions for Simulation 1. The dotted lines represent the marginal location parameters with the orange as the marginal location for group 1 and the yellow for group 2.}
\label{fig:Sim1}
\end{figure}
\begin{figure}
\centering
\includegraphics[width=0.85\textwidth]{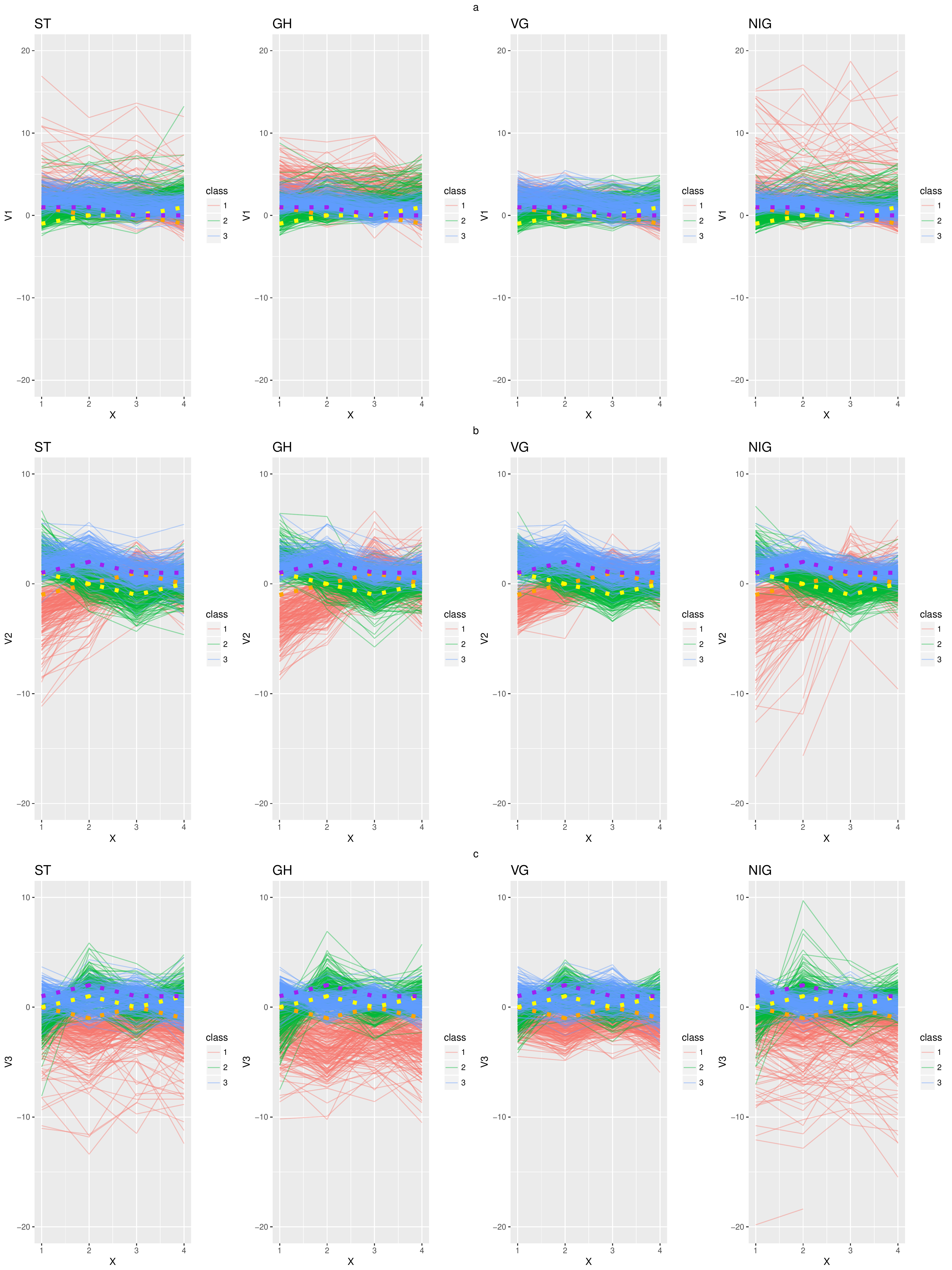}
\caption{Marginal data for the columns for each of the four distributions for Simulation 2. The dotted lines represent the marginal location parameters with the orange as the marginal location for group 1, yellow for group 2, and purple for group 3.}
\label{fig:Sim2}
\end{figure}

\end{document}